\documentclass[10pt,twocolumn,amsmath,amssymb,aps,pra,showpacs,longbibliography,superscriptaddress]{revtex4-1}
\usepackage[latin9]{inputenc}
\usepackage{amsmath}
\usepackage{amssymb}
\usepackage{graphicx}
\usepackage{wasysym}
\usepackage{esint}

\makeatletter
\usepackage{amsfonts}
\usepackage{graphics}

\makeatother

\begin{document}
\title{Altermagnetism-driven FFLO superconductivity in finite-filling 2D
lattices}
\author{Xia-Ji Liu}
\email{Corresponding author: xiajiliu@swin.edu.au}

\affiliation{Centre for Quantum Technology Theory, Swinburne University of Technology,
Melbourne 3122, Australia}
\author{Hui Hu}
\affiliation{Centre for Quantum Technology Theory, Swinburne University of Technology,
Melbourne 3122, Australia}
\date{\today}
\begin{abstract}
We systematically investigate the emergence of finite-momentum Fulde-Ferrell-Larkin-Ovchinnikov
(FFLO) superconductivity in a square lattice Hubbard model with finite
filling, driven by either $d_{xy}$-wave or $d_{x^{2}-y^{2}}$-wave
altermagnetic order in the presence of on-site $s$-wave attractive
interactions. Our study combines mean-field calculation in the superconducting
phase with pairing instability analysis of the normal state, incorporating
the next-nearest-neighbor hopping in the single-particle dispersion
relation. We demonstrate that the two types of altermagnetism have
markedly different impacts on the stabilization of FFLO states. Specifically,
$d_{xy}$-wave altermagnetism supports FFLO superconductivity over
a broad parameter regime at low fillings, whereas $d_{x^{2}-y^{2}}$-wave
altermagnetism only induces FFLO pairing in a narrow range at high
fillings. Furthermore, we find that the presence of a Van Hove singularity
in the density of states tends to suppress FFLO superconductivity.
These findings may provide guidance for experimental exploration of
altermagnetism-induced FFLO states in real materials with more complex
electronic structures.
\end{abstract}
\maketitle

\section{Introduction}

Altermagnetism is a newly-identified type of magnetic ordering in
crystalline materials \citep{Smejkal2022,Jungwirth2025,Liu2025Review},
distinct from the two long\nobreakdash-known classes of ferromagnetism
and antiferromagnetism. In altermagnetic materials (commonly called
altermagnets), despite the total magnetic moment adding up to zero
(as in antiferromagnets), the electronic bands exhibit a significant
spin splitting (a trait of ferromagnets) \citep{Noda2016,Naka2019,Hayami2019,Ahn2019,Hayami2020a,Hayami2020b,Smejkal2020,Mazin2023,Amin2024}.
This unique combination makes altermagnets compelling for ultrafast
spintronics applications with improved scalability. The presence of
spin\nobreakdash-splitting is especially intriguing and opens up a
new frontier in exploring novel quantum many-body phenomena in condensed
matter physics \citep{Jungwirth2025,Liu2025Review}.

In this context, the prospect of unconventional superconductivity
driven by altermagnetism has garnered increasing attention \citep{Liu2025Review,deCarvalho2024,Mazin2022,Fukaya2025a}.
In particular, Zhang, Hu and Neupert predicted the possibility of
finite\nobreakdash-momentum Cooper pairing by modeling altermagnets
in proximity to conventional $s$\nobreakdash-wave superconductors
\citep{Zhang2024}. That intriguing proposal opens the way for spatially
inhomogeneous Fulde--Ferrell--Larkin--Ovchinnikov (FFLO) state
superconductivity \citep{Fulde1964,Larkin1964,Casalbuoni2004,Uji2006,Hu2006,Hu2007,Kenzelmann2008,Liu2013,Wan2023,Zhao2023,Kawamura2024}
in altermagnetic metals \citep{Sumita2023,Chakraborty2024,Hong2025,Iorsh2025,Sim2024,Sumita2025,Hu2025PRB,Mukasa2025,Liu2025PRB}.
Indeed, earlier work on quantum spin\nobreakdash-nematic phases established
that a $d$\nobreakdash-wave spin\nobreakdash-splitting band structure
can give rise to FFLO states in the presence of attractive $d$\nobreakdash-wave
interactions \citep{Soto-Garrido2014}. More recently, Chakraborty
and Black\nobreakdash-Schaffer confirmed the emergence of FFLO superconductivity
induced by $d$\nobreakdash-wave altermagnetism with $d$\nobreakdash-wave
interaction potentials \citep{Chakraborty2024}. A subsequent study
by Hong, Park and Kim \citep{Hong2025} demonstrated that such an
FFLO state can also arise under $s$\nobreakdash-wave interactions
- challenging the earlier belief that $d$\nobreakdash-wave interactions
are essential. That discrepancy has now been resolved through further
analytic and numerical work by Hu and collaborators \citep{Hu2025PRB,Liu2025PRB}.

It is worth noting that most of the aforementioned theoretical studies
on FFLO superconductivity adopted a single-band model of altermagnetism.
More realistic modeling that incorporates the sub-lattice degree of
freedom was performed by Sumita, Naka and Seo \citep{Sumita2025}.
Based on a Ginzburg--Landau framework, their detailed analysis clearly
showed that a phase\nobreakdash-modulated (plane\nobreakdash-wave-like)
Fulde--Ferrell form of the superconducting order parameter is favored
when a two-band model for sub\nobreakdash-lattices are used, rather
than the amplitude\nobreakdash-modulated Larkin\nobreakdash-Ovchinnikov
type or other more complex FFLO configurations \citep{Sumita2025}.

In this work, we would like to revisit altermagnetism-driven FFLO
superfluidity in two-dimensional lattices by systematically examining
various types of $d$-wave altermagnetism using more realistic single-particle
dispersion relations at different electron filling factors. The motivation
for this study is two-fold.

First, previous research has primarily focused on either $d_{x^{2}-y^{2}}$\nobreakdash-wave
\citep{Chakraborty2024} or $d_{xy}$\nobreakdash-wave altermagnetism
\citep{Hong2025,Hu2025PRB}. While these two types are equivalent
in the dilute limit with very low filling, they may exhibit distinct
effects on superconducting states at finite or high filling factors.
A comparative investigation is therefore essential to clarify their
differing roles in inducing FFLO superconductivity. On the other hand,
most earlier studies assumed a single-particle dispersion including
only the nearest-neighbor hopping. It is thus of interest to explore
altermagnetism-driven FFLO superconductivity within more complex and
realistic electronic structures, for instance, when next-nearest-neighbor
hopping is present. Such modification shifts the Van Hove singularity
in the electronic density of states away from half-filling, raising
questions about how this singularity influences superconducting behavior
\citep{Sumita2025}.

We employ two complementary theoretical approaches to investigate
FFLO superconductivity. In the superconducting regime, we explicitly
solve the mean-field Bogoliubov--de Gennes equations to determine
the finite-momentum order parameter. In the normal state with relatively
strong altermagnetic coupling, we analyze the onset of finite-momentum
pairing instability using the well-established Thouless criterion.
Together, these approaches provide a comprehensive phase diagram of
the superconducting states as functions of the filling factor and
altermagnetic coupling strength.

The remainder of this paper is organized as follows. In Sec. II, we
introduce the model Hamiltonian that describes $d$-wave altermagnetic
metals in two-dimensional lattices. Section III outlines the solution
of this model using standard mean-field approaches. In Sec. IV, we
construct and analyze the phase diagram for both $d_{x^{2}-y^{2}}$\nobreakdash-wave
and $d_{xy}$\nobreakdash-wave altermagnetism. Finally, Sec. V provides
concluding remarks. In Appendix A, we consider the temperature effect.
In Appendix B, we present additional results at low filling factors,
explicitly demonstrating the equivalence between $d_{x^{2}-y^{2}}$\nobreakdash-wave
and $d_{xy}$\nobreakdash-wave altermagnetism in driving FFLO superconductivity
in the dilute limit.

\begin{figure}[t]
\begin{centering}
\includegraphics[width=0.5\textwidth]{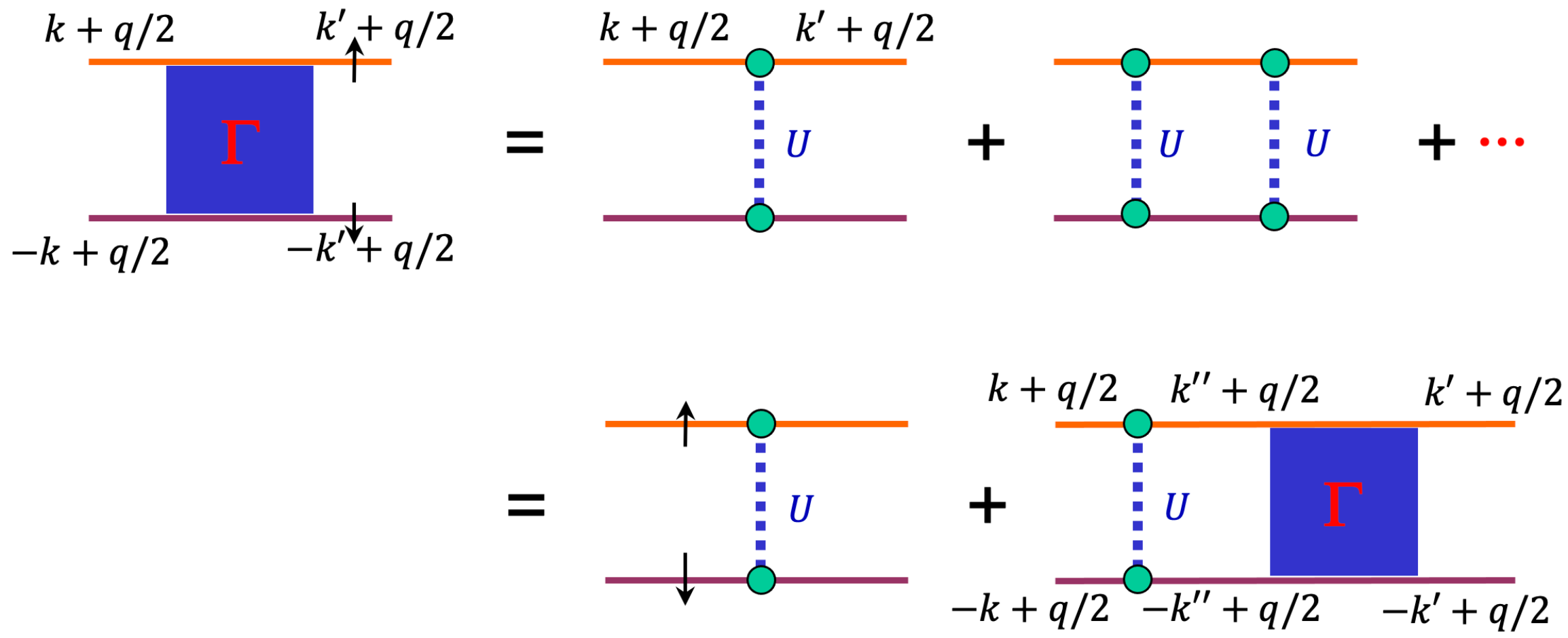}
\par\end{centering}
\caption{\label{fig1: VertexFunction} Diagrammatic representation of the two-particle
vertex function $\Gamma(\mathbf{k},\mathbf{k}';\mathbf{q},\omega)$,
within the ladder approximation.}
\end{figure}

\section{Model Hamiltonian}

We describe a 2D altermagnetic metal on square lattices with an area
$\mathcal{S}$, by using the effective single-band Hamiltonian, $\mathcal{H}=\mathcal{H}_{0}+\mathcal{H}_{\text{int}}$,
where 
\begin{eqnarray}
\mathcal{H}_{0} & = & \sum_{\mathbf{k}\sigma}\left[\xi_{\mathbf{k}}+s\left(\sigma\right)J_{\mathbf{k}}\right]c_{\mathbf{k}\sigma}^{\dagger}c_{\mathbf{k}\sigma},\\
\mathcal{H}_{\textrm{int}} & = & \frac{U}{\mathcal{S}}\sum_{\mathbf{k},\mathbf{k}',\mathbf{q}}c_{\mathbf{k}+\frac{\mathbf{q}}{2}\uparrow}^{\dagger}c_{\mathbf{-k}+\frac{\mathbf{q}}{2}\downarrow}^{\dagger}c_{\mathbf{-k'}+\frac{\mathbf{q}}{2}\downarrow}c_{\mathbf{k}'+\frac{\mathbf{q}}{2}\uparrow}
\end{eqnarray}
are the non-interacting Hamiltonian and the interacting Hamiltonian,
respectively, and $c_{\mathbf{k}\sigma}^{\dagger}$ ($c_{\mathbf{k}\sigma}$)
are the creation (annihilation) field operators. We consider the spin-independent
single-particle dispersion relation \citep{Romer2015},
\begin{equation}
\xi_{\mathbf{k}}=-2t(\cos k_{x}+\cos k_{y})-4t'\cos k_{x}\cos k_{y}-\mu,
\end{equation}
where $t$ and $t'$ are the nearest-neighbor and the next-to-nearest-neighbor
hopping strengths, and $\mu$ is the chemical potential that controls
the lattice filling factor. The presence of the altermagnetism introduces
a spin-dependent band-splitting, $s(\uparrow)J_{\mathbf{k}}=+J_{\mathbf{k}}$
and $s(\downarrow)J_{\mathbf{k}}=-J_{\mathbf{k}}$, for spin-up and
spin-down electrons, respectively. We include the two forms of the
$d$-wave altermagnetism,
\begin{equation}
J_{\mathbf{k}}=-\lambda\sin k_{x}\sin k_{y}
\end{equation}
for the $d_{xy}$-wave type \citep{Hong2025} and 

\begin{equation}
J_{\mathbf{k}}=-\frac{t_{\textrm{am}}}{2}\left(\cos k_{x}-\cos k_{y}\right)
\end{equation}
for the $d_{x^{2}-y^{2}}$-wave type \citep{Chakraborty2024}, respectively.
For simplicity, here we consider the on-site attraction $U<0$ only,
which gives rise to the spin-singlet pairing between two electrons
with unlike spins.

\section{Theoretical framework}

We now present the theoretical approaches used to study the FFLO state,
both when the system is in the normal (non-superconducting) phase,
and when it is already superconducting.

\subsection{Thouless criterion for the pairing instability}

In the normal phase, we examine whether the system becomes unstable
to pair formation for given parameters, such as temperature $T$,
interaction strength $U$, and the altermagnetic coupling constant
$\lambda$ or $t_{\textrm{am}}$. A convenient way to check this is
via the Thouless criterion, which states that the instability sets
in when the inverse two-particle vertex function satisfies \citep{Thouless1960,Liu2006},
\begin{equation}
\max_{\{\mathbf{q}\}}\Gamma^{-1}\left(\mathbf{q},\omega=0\right)=0.\label{eq:ThoulessCriterion}
\end{equation}
In the case that the maximum of the inverse vertex function occurs
at a nonzero momentum $\mathbf{q}\neq0$, the normal state is unstable
towards the formation of an inhomogeneous finite-momentum FFLO state.

In this work, we use the standard non-self-consistent $T$-matrix
approach \citep{Thouless1960,Liu2006}. By summing all the ladder
diagrams illustrated in Fig. \ref{fig1: VertexFunction}, we may formally
write the equation satisfied by the approximate vertex function,\begin{widetext}
\begin{equation}
\Gamma\left(\mathbf{k},\mathbf{k}';\mathbf{q},\omega\right)=U-\sum_{k''}UG_{0\uparrow}\left(k''+\frac{q}{2}\right)G_{0\downarrow}\left(-k''+\frac{q}{2}\right)\Gamma\left(\mathbf{k}'',\mathbf{k}';\mathbf{q},\omega\right),
\end{equation}
where we have introduced the four-momenta $q=\{\mathbf{q},i\nu_{n}\rightarrow\omega+i0^{+}\}$
and $k''=\{\mathbf{k}'',i\omega_{m}\}$, with the bosonic and fermionic
Matsubara frequencies $\nu_{n}=2\pi nT$ and $\omega_{m}=2\pi(m+1/2)T$
($n,m\in\mathbb{Z}$), respectively. We have also abbreviated the
summation $\sum_{k''}\equiv(k_{B}T/\mathcal{S})\sum_{\omega_{m}}\sum_{\mathbf{k}''}$.
It is readily seen that the two-particle vertex function is independent
on $\mathbf{k}$ and $\mathbf{k'}$, due to the contact-like on-site
interaction $U$. The summation over the fermionic Matsubara frequency
$\omega_{m}$ can be easily carried out, and we obtain,
\begin{equation}
\Gamma^{-1}\left(\mathbf{q},\omega\right)=\frac{1}{U}+\frac{1}{\mathcal{S}}\sum_{\mathbf{k}''}\frac{\left[n_{F}\left(\xi_{\frac{\mathbf{q}}{2}+\mathbf{k''}\uparrow}\right)+n_{F}\left(\xi_{\frac{\mathbf{q}}{2}-\mathbf{k''}\uparrow}\right)-1\right]}{\omega^{+}-\left(\xi_{\frac{\mathbf{q}}{2}+\mathbf{k''}\uparrow}+\xi_{\frac{\mathbf{q}}{2}-\mathbf{k}''\uparrow}\right)},\label{eq:InvGamma}
\end{equation}
\end{widetext}where $\xi_{\mathbf{k}\uparrow}=\xi_{\mathbf{k}}+J_{\mathbf{k}}$
and $\xi_{\mathbf{k}\downarrow}=\xi_{\mathbf{k}}-J_{\mathbf{k}}$,
and $n_{F}(x)\equiv1/[e^{x/(k_{B}T)}+1]$ is the Fermi-Dirac distribution
function.

We perform a numerical integration of the momentum $\mathbf{k''}$
in Eq. (\ref{eq:InvGamma}) across the first Brillouin zone of the
square lattice, using a small imaginary broadening $\eta\sim0.02t$.
The resulting numerical precision of the inverse vertex function can
be roughly estimated by verifying that the identity 
\begin{equation}
\textrm{Im}\Gamma^{-1}\left(\mathbf{q},\omega=0\right)=0
\end{equation}
 is satisfied within some tolerance.

\begin{figure}
\begin{centering}
\includegraphics[width=0.45\textwidth]{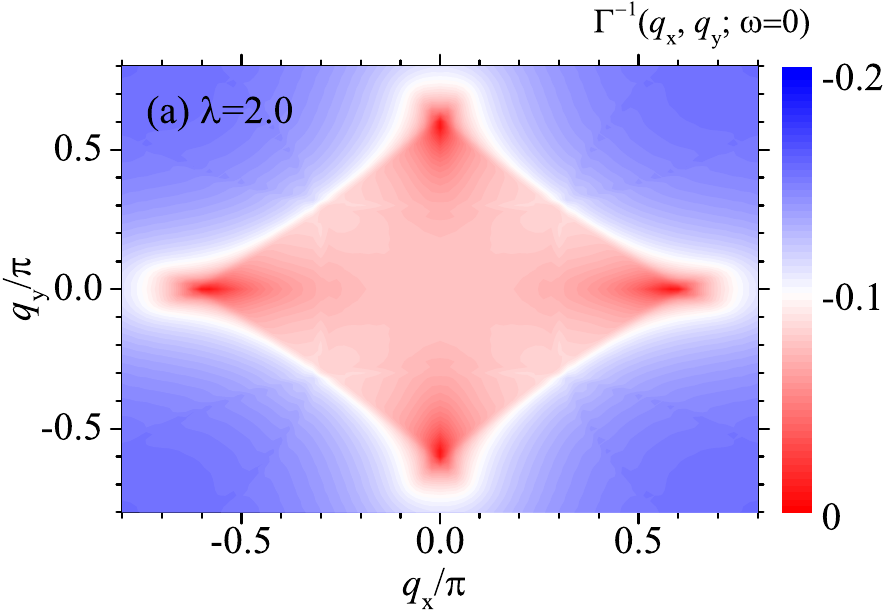}
\par\end{centering}
\begin{centering}
\includegraphics[width=0.45\textwidth]{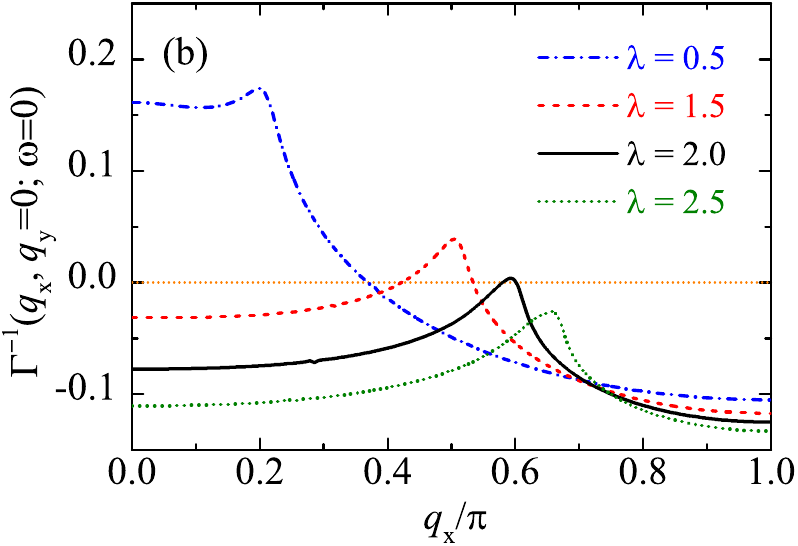}
\par\end{centering}
\caption{\label{fig2:dxy_thouless} (a) The inverse vertex function at zero
frequency $\Gamma^{-1}(\mathbf{q},\omega=0)$, in units of $t^{-1}$,
in the $q_{x}$-$q_{y}$ plane at the altermagnetic coupling $\lambda=2t$.
The inverse vertex function reaches maximum at $(q_{x},q_{y})=(\pm q,0)$
or $(0,\pm q)$, where $q=\left|\mathbf{q}\right|$. (b) $\Gamma^{-1}(q_{x},q_{y}=0;\omega=0)$
at different altermagnetic couplings $\lambda$ as indicated. Here,
we consider the $d_{xy}$-wave altermagnetism and take an attraction
strength $U=-3t$ at the filling factor $\nu=1.0$. We have taken
a negligible temperature $T=0.01t$ to smooth the sharp Fermi surface
and to increase numerical accuracy.}
\end{figure}

\begin{figure*}
\begin{centering}
\includegraphics[width=0.33\textwidth]{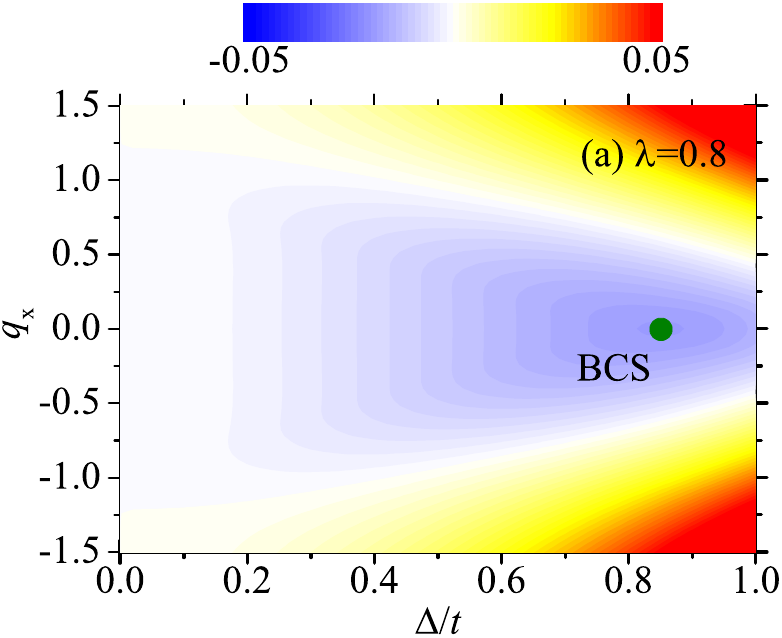}\includegraphics[width=0.33\textwidth]{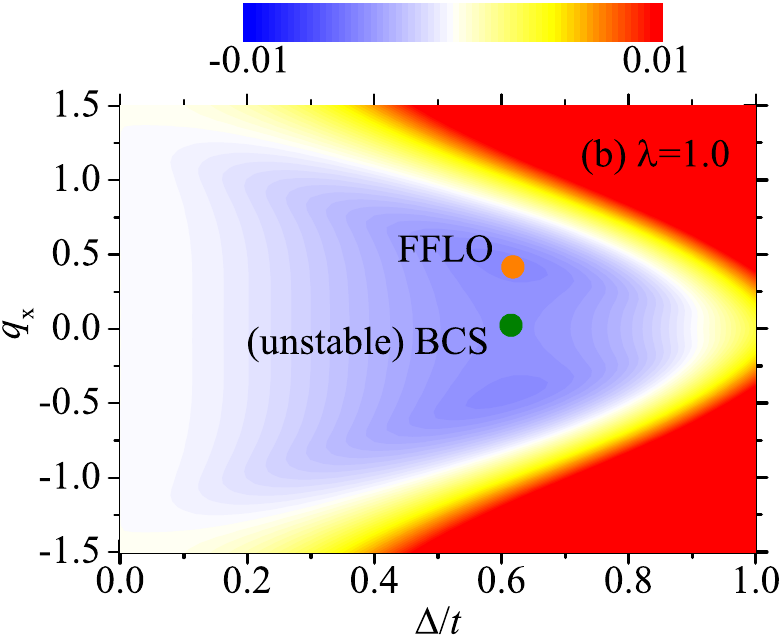}\includegraphics[width=0.33\textwidth]{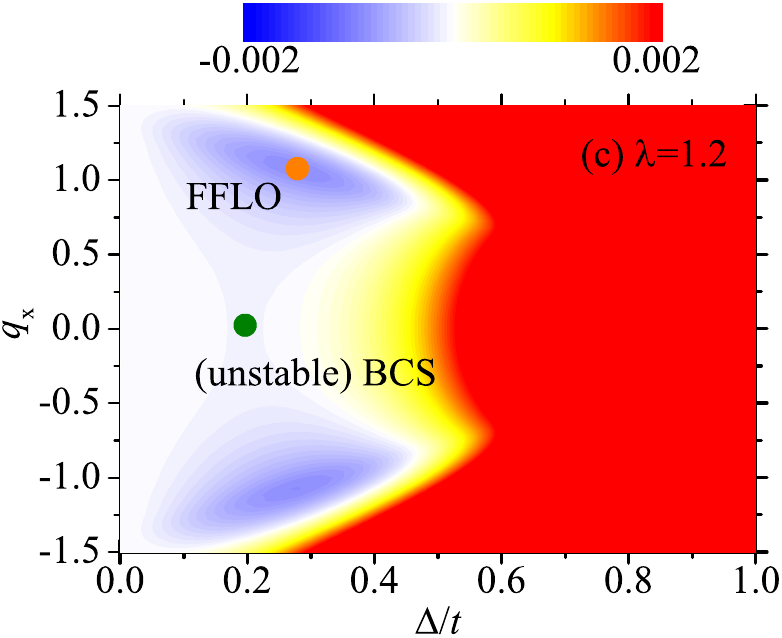}
\par\end{centering}
\caption{\label{fig3: Omega2D} The contour plots of the thermodynamic potential
$(\Omega-\Omega_{0})/(\nu t)$, as functions of the pairing amplitude
$\Delta$ and the FFLO momentum $q_{x}$, at three altermagnetic coupling
strengths: $\lambda=0.8t$ (a), $\lambda=1.0t$ (b) and $\lambda=1.2t$
(c). The chemical potential in each plot is, $\mu=2.05t$ (a), $\mu=1.98t$
(b) and $\mu=1.91t$ (c). The green dots indicate the stable or unstable
BCS states, while the orange dots highlight the FFLO state. Here,
we consider the $d_{xy}$-wave altermagnetism and take an attraction
strength $U=-3t$ at the filling factor $\nu=1.0$.}
\end{figure*}

\subsection{Mean-field theory}

In the superconducting phase, we assume a (real) order parameter $\Delta=-(U/\mathcal{S})\sum_{\mathbf{k}}\left\langle c_{\mathbf{-k}+\mathbf{q}/2\downarrow}c_{\mathbf{k}+\mathbf{q}/2\uparrow}\right\rangle $
in momentum space to search for the inhomogeneous superconductivity
at a finite-momentum $\mathbf{q}$ \citep{Hu2006,Hu2025PRB}, whose
direction is determined by the type of altermagnetism. In real space,
this order parameter corresponds to a phase-modulated form, $\Delta(\mathbf{x})=\Delta e^{i\mathbf{q}\cdot\mathbf{x}}$.
By taking the mean-field decoupling of the interaction Hamiltonian,
\begin{equation}
{\cal H}_{\textrm{int}}\simeq-\sum_{\mathbf{k}}\left[\Delta c_{\mathbf{k}+\frac{\mathbf{q}}{2}\uparrow}^{\dagger}c_{\mathbf{-k}+\frac{\mathbf{q}}{2}\downarrow}^{\dagger}+\text{H.c.}\right]-\frac{\Delta^{2}}{U}\mathcal{S}.
\end{equation}
we end up with a bilinear pairing Hamiltonian, 
\begin{eqnarray}
{\cal H} & = & \sum_{\mathbf{k}}\left[\xi_{\mathbf{k}+\frac{\mathbf{q}}{2}\uparrow}c_{\mathbf{k}+\frac{\mathbf{q}}{2}\uparrow}^{\dagger}c_{\mathbf{k}+\frac{\mathbf{q}}{2}\uparrow}+\xi_{\frac{\mathbf{q}}{2}-\mathbf{k}\downarrow}c_{\mathbf{-k}+\frac{\mathbf{q}}{2}\downarrow}^{\dagger}c_{-\mathbf{k}+\frac{\mathbf{q}}{2}\downarrow}\right]\nonumber \\
 &  & -\sum_{\mathbf{k}}\left[\Delta c_{\mathbf{k}+\frac{\mathbf{q}}{2}\uparrow}^{\dagger}c_{\mathbf{-k}+\frac{\mathbf{q}}{2}\downarrow}^{\dagger}+\text{H.c.}\right]-\frac{\Delta^{2}}{U}\mathcal{S},
\end{eqnarray}
which can be solved by using the standard Bogoliubov transformation,
with the introduction of the Nambu spinor,
\begin{equation}
\Psi_{\mathbf{k}}=\left[\begin{array}{c}
c_{\mathbf{k}+\frac{\mathbf{q}}{2}\uparrow}\\
c_{\mathbf{-k}+\frac{\mathbf{q}}{2}\downarrow}^{\dagger}
\end{array}\right].
\end{equation}
More explicitly, we rewrite the pairing Hamiltonian in the form,
\begin{equation}
\mathcal{H}=\sum_{\mathbf{k}}\Psi_{\mathbf{k}}^{\dagger}\left[\begin{array}{cc}
\xi_{\mathbf{k}+\frac{\mathbf{q}}{2}\uparrow} & -\Delta\\
-\Delta & \xi_{\frac{\mathbf{q}}{2}-\mathbf{k}\downarrow}
\end{array}\right]\Psi_{\mathbf{k}}-\frac{\Delta^{2}}{U}\mathcal{S}+\sum_{\mathbf{k}}\xi_{\frac{\mathbf{q}}{2}-\mathbf{k}\downarrow}.
\end{equation}
The diagonalization of the two by two matrix leads to two Bogoliubov
excitation spectra \citep{Hong2025}, indexed by $\eta=\pm$:
\begin{equation}
E_{\mathbf{k}\pm}=\sqrt{A_{\mathbf{k}}^{2}+\Delta^{2}}\pm B_{\mathbf{k}},
\end{equation}
where 
\begin{eqnarray}
A_{\mathbf{k}} & \equiv & \frac{1}{2}\left(\xi_{\frac{\mathbf{q}}{2}+\mathbf{k}}+\xi_{\frac{\mathbf{q}}{2}-\mathbf{k}}+J_{\frac{\mathbf{q}}{2}+\mathbf{k}}-J_{\frac{\mathbf{q}}{2}-\mathbf{k}}\right),\\
B_{\mathbf{k}} & \equiv & \frac{1}{2}\left(\xi_{\frac{\mathbf{q}}{2}+\mathbf{k}}-\xi_{\frac{\mathbf{q}}{2}-\mathbf{k}}+J_{\frac{\mathbf{q}}{2}+\mathbf{k}}+J_{\frac{\mathbf{q}}{2}-\mathbf{k}}\right).
\end{eqnarray}
The corresponding grand thermodynamic potential at finite temperature
$T$ takes the form,
\begin{equation}
\Omega=-\frac{\Delta^{2}}{U}\mathcal{S}+\mathcal{E}_{0}-k_{B}T\sum_{{\bf k},\eta=\pm}\ln\left(1+e^{-\frac{E_{{\bf k\eta}}}{k_{B}T}}\right),
\end{equation}
where 
\begin{equation}
\mathcal{E}_{0}\equiv\sum_{\mathbf{k}}\left(\xi_{\frac{\mathbf{q}}{2}-\mathbf{k}\downarrow}-E_{\mathbf{k}-}\right)=\sum_{\mathbf{k}}\left(A_{\mathbf{k}}-\sqrt{A_{\mathbf{k}}^{2}+\Delta^{2}}\right)
\end{equation}
denotes the energy offset that arises from performing the Bogoliubov
transformation - in the process of expressing the quasiparticle field
operators in proper order, and the last term is the usual free-fermion
contribution from Bogoliubov quasiparticles. For fixed chemical potential
$\mu$ (measured from the bottom of the single-particle dispersion
relation $\xi_{\mathbf{k}}$) and temperature $T$, the two independent
parameters are the order parameter $\Delta$ and the FFLO momentum
$q$. Their equilibrium values are determined by minimizing the thermodynamic
potential, i.e., 
\begin{eqnarray}
\frac{\partial\Omega}{\partial\Delta} & = & 0,\\
\frac{\partial\Omega}{\partial q} & = & 0.
\end{eqnarray}
Finally, the chemical potential is adjusted so that the number equation,
\begin{equation}
\nu=-\partial\Omega/\partial\mu,
\end{equation}
is satisfied for the given filling factor $\nu$.

\section{Results and discussions}

Throughout this work, we take the next-to-nearest-neighbor hopping
strength $t'=-0.35t$ \citep{Romer2015}, as a typical value for the
candidate altermagnetic materials. This nonzero hopping strength $t'$
breaks the particle-hole symmetry at the half-filling $\nu=1$ and
moves the position of the Van Hove singularity to a smaller filling
factor $\nu_{\textrm{VH}}\simeq0.66$ \citep{Romer2015}. We also
take an on-site attraction $U=-3t$ and an essentially zero temperature
$T=0.01t$, unless specified otherwise. In the following, we discuss
separately the two cases of the $d_{xy}$- and $d_{x^{2}-y^{2}}$-wave
altermagnetism.

\subsection{$d_{xy}$-wave altermagnetism}

In Fig. \ref{fig2:dxy_thouless}(a), we report a contour map of the
zero-frequency inverse vertex function $\Gamma^{-1}(\mathbf{q},\omega=0)$
as a function of the pairing momentum components $q_{x}$ and $q_{y}$,
for $d_{xy}$-wave altermagnetism with coupling strength $\lambda=2t$
at half-filling ($\nu=1$). It is immediately evident that the maximum
of $\Gamma^{-1}$ lies along either the $q_{x}$-axis or the $q_{y}$-axis
\citep{Hong2025}. Importantly, this behavior remains robust across
other filling factors $\nu$ and different altermagnetic coupling
constants $\lambda$. As a consequence, in our numerical calculations
we may without loss of generality take $\mathbf{q}=q_{x}\mathbf{e}_{x}$
for $d_{xy}$-wave altermagnetism. Then, in Fig. \ref{fig2:dxy_thouless}(b)
we plot the dependence of $\Gamma^{-1}(\mathbf{q},\omega=0)$ on $q_{x}$
for several values of $\lambda$. According to the Thouless criterion,
a non-negative value of $\Gamma^{-1}$ signals that the system enters
a superconducting phase. We observe that the transition to this superconducting
phase sets in around $\lambda\simeq2t$, where the nonzero optimal
pairing momentum $q_{x,\textrm{max}}\simeq0.6\pi$ indicates that
the emergent superconducting state is spatially inhomogeneous.

To analyze the symmetry-breaking phases in the superconducting domain,
we apply the Bogoliubov mean-field approach. In Fig. \ref{fig3: Omega2D},
we show the contour plots of the thermodynamic potential $\Omega$
versus the pairing amplitude $\Delta$ and the FF momentum $q_{x}$,
for various values of the altermagnetic coupling strengths, taking
the non-interacting thermodynamic potential $\Omega_{0}$ as a reference
point. For weak altermagnetic coupling (see Fig. \ref{fig3: Omega2D}(a)),
the potential landscape exhibits a single global minimum, corresponding
to the conventional uniform BCS state with $q_{x}=0$. As the altermagnetic
coupling strength grows (Fig. \ref{fig3: Omega2D}(b)), the BCS minimum
becomes unstable (i.e., turning into a saddle point), while a new
global minimum emerges at nonzero $\pm q_{x}$, indicating the onset
of an inhomogeneous FFLO phase. By further increasing $\lambda$ (Fig.
\ref{fig3: Omega2D}(c)), the FFLO minimum moves well aways from the
now-unstable BCS saddle point and becomes increasingly pronounced.

\begin{figure}
\begin{centering}
\includegraphics[width=0.5\textwidth]{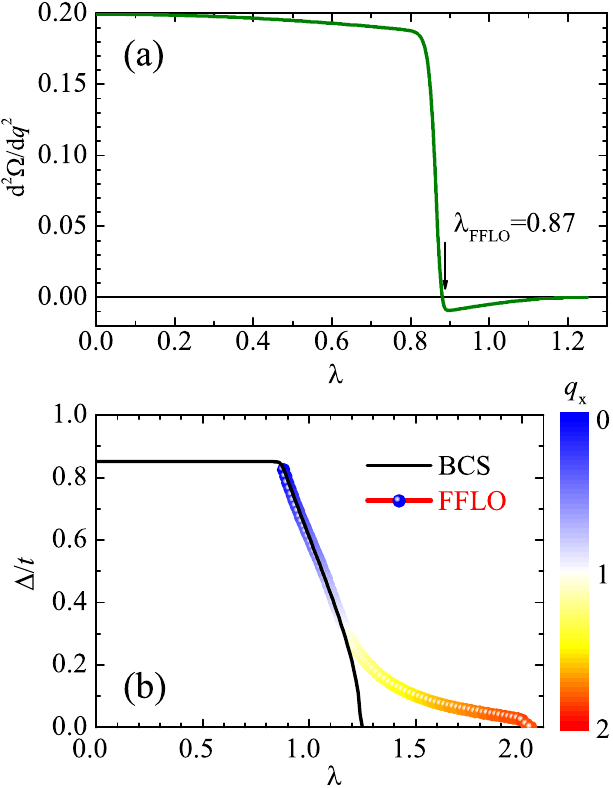}
\par\end{centering}
\caption{\label{fig4: nu100} (a) The second derivative $\partial^{2}\Omega/\partial q_{x}^{2}$
of the BCS state as a function of $\lambda$. It becomes negative
at $\lambda_{\textrm{FFLO}}\simeq0.87t$, where the FFLO state starts
to emerge. (b) The pairing gaps of the BCS state (solid line) and
of the FFLO state (colored dots) as a function of the altermagnetic
coupling strength $\lambda$. The color of the dots shows the values
of the FF momentum $q_{x}$, following the color map. Here, we consider
the $d_{xy}$-wave altermagnetism and take an attraction strength
$U=-3t$ at the filling factor $\nu=1.0$. }
\end{figure}

The above instability of the uniform BCS state toward an inhomogeneous
FFLO phase could be tracked by the phase stiffness: a negative stiffness
under a phase twist (i.e., $\rho_{s}\propto\partial\Omega^{2}/\partial q^{2}<0$)
signals that the uniform state is unstable to finite-momentum pairing.
In Fig. \ref{fig4: nu100}(a), we present the second derivative $\partial^{2}\Omega/\partial q_{x}^{2}$
with increasing the altermagnetic coupling. We find that this curvature
decreases sharply once $\lambda>0.8t$ and becomes negative around
a critical coupling $\lambda_{\textrm{FFLO}}\simeq0.87$, indicating
the loss of stability of the uniform phase. Beyond this threshold,
the pairing amplitude of the (meta-stable) BCS solution drops quickly
and vanishes near $\lambda\sim1.25t$, as seen in Fig. \ref{fig4: nu100}(b).
The stable FFLO phase has a relatively small pairing amplitude, but
it persists up to $\lambda_{c}\sim2t$, thus providing a relative
wide parameter range in which the FFLO state appears.

\begin{figure}
\begin{centering}
\includegraphics[width=0.5\textwidth]{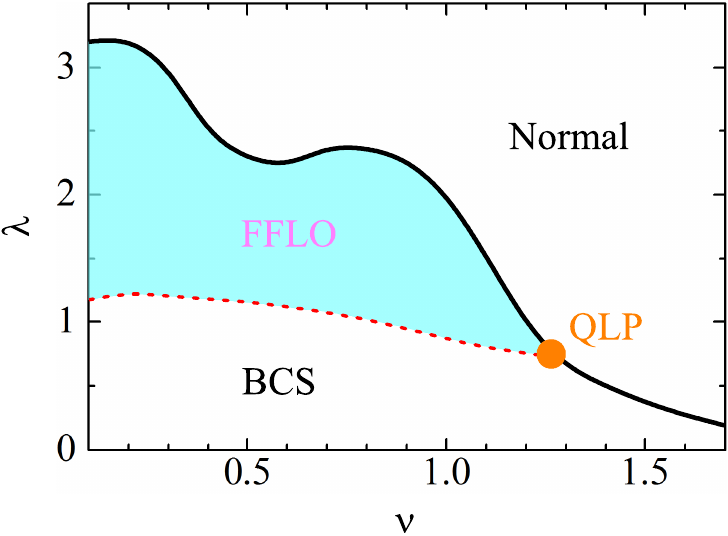}
\par\end{centering}
\caption{\label{fig5: dxyPhaseDiagram} The phase diagram under the $d_{xy}$-wave
altermagnetism, as functions of the filling factor $\nu$ and the
altermagnetic coupling strength $\lambda$. The transition from the
BCS state to the FFLO state is shown by the red dashed line, while
the transition from the FFLO state to the normal state is shown by
the black solid line. The three phases meet at a quantum Lifshitz
point, as highlighted by an orange dot. Here, we take an attraction
strength $U=-3t$. }
\end{figure}

We repeat the numerical analysis at different filling factors to construct
a superconducting phase diagram under $d_{xy}$-wave altermagnetism,
as reported in Fig. \ref{fig5: dxyPhaseDiagram}. In this diagram,
the boundary marking the transition from the uniform BCS state to
the FFLO phase ($\lambda_{\textrm{FFLO}}$) is indicated by the red
dashed line, and the line marking the transition from the FFLO phase
to the normal state ($\lambda_{c}$) is shown as a solid black curve.
As the filling factor increases, the threshold coupling $\lambda_{\textrm{FFLO}}$
steadily decreases, while the critical coupling $\lambda_{c}$ exhibits
a non-monotonic dependence on filling. Near $\nu\simeq0.6$, the parameter
range supporting the FFLO phase narrows significantly, a feature that
might be associated with proximity to the Van Hove singularity in
the density of states. The FFLO window eventually disappears at large
filling (i.e., $\nu\simeq1.25$). As a result, the phase diagram reveals
a quantum Lifshitz point at which the BCS, FFLO, and normal phases
converge, marking a multi-critical point where all three phases meet
\citep{Pisarski2019,Hu2025AB}.

\begin{figure}
\begin{centering}
\includegraphics[width=0.5\textwidth]{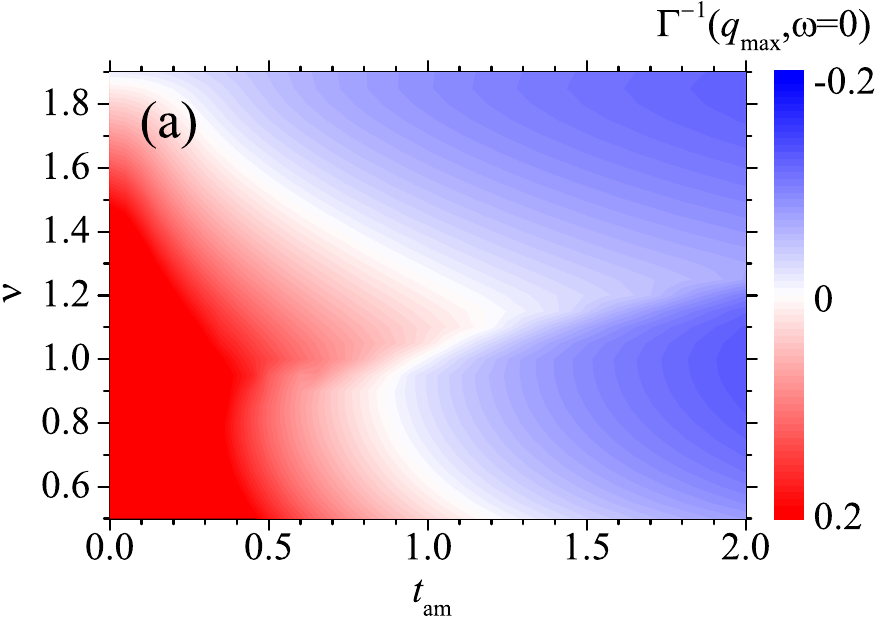}
\par\end{centering}
\begin{centering}
\includegraphics[width=0.45\textwidth]{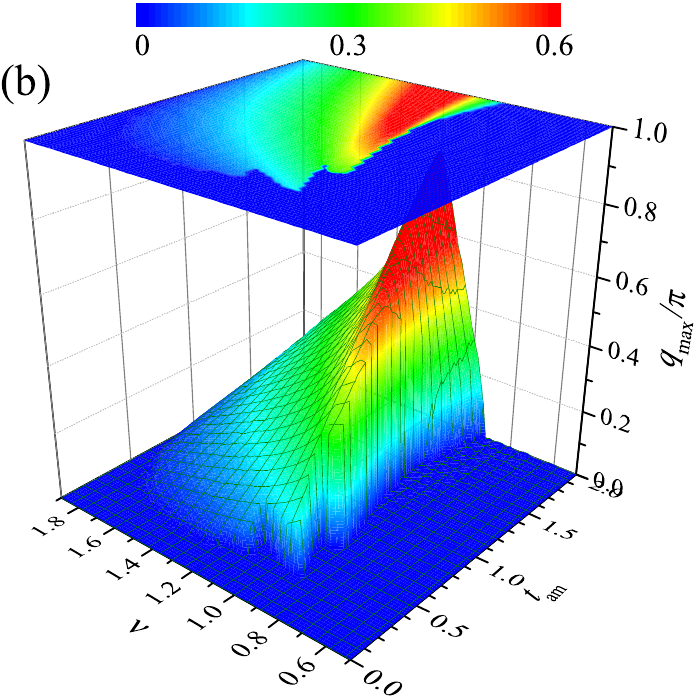}
\par\end{centering}
\caption{\label{fig6: dx2y2_thouless} (a) The maximum of the inverse vertex
function $\Gamma^{-1}(q_{\max},\omega=0)$ and the corresponding pairing
momentum $q_{\max}$ (b) as functions of the altermagnetic coupling
$t_{\textrm{am}}$ and the filling factor $\nu$. The contour line
in white color in (a) indicates the Thouless criterion $\Gamma^{-1}=0$
that separates the superconducting phase from the normal state. Here,
we consider the $d_{x^{2}-y^{2}}$-wave altermagnetism and take an
attraction strength $U=-3t$. In this case, the FFLO pairing instability
occurs at the filling factor $\nu\apprge1$ only.}
\end{figure}

\subsection{$d_{x^{2}-y^{2}}$-wave altermagnetism}

We now examine superconductivity in the presence of $d_{x^{2}-y^{2}}$-wave
altermagnetism. In this situation, the inverse vertex function $\Gamma^{-1}(\mathbf{q},\omega=0)$
attains its maximum along the diagonal (or off-diagonal) directions,
namely at $q_{x}=\pm q_{\textrm{max}}/\sqrt{2}$ and $q_{y}=\pm q_{\textrm{max}}/\sqrt{2}$.
Consequently, in our numerical analysis we take $(q_{x},q_{y})=(q,q)/\sqrt{2}$
without loss of generality.

In Fig. \ref{fig6: dx2y2_thouless}, we present the contour maps of
$\Gamma^{-1}(q=q_{\textrm{max}},\omega=0)$ and of $q_{\textrm{max}}$
as a function of the $d_{x^{2}-y^{2}}$-wave altermagnetic coupling
$t_{\textrm{am}}$ and the filling factor $\nu$. According to the
Thouless criterion in Eq. (\ref{eq:ThoulessCriterion}), the superconducting
phase (highlighted in red) is enclosed by the white contour line (see
Fig. \ref{fig6: dx2y2_thouless}(a)). This boundary exhibits a pronounced,
non-monotonic dependence on the filling factor. As before, we observe
a suppression of the superconducting region around $\nu\sim0.8$,
which might be related to the appearance of the Van Hove singularity.
Remarkably, from Fig. \ref{fig6: dx2y2_thouless}(b) we find that
a finite-momentum pairing emerges only at sufficiently large filling
factors. As the filling factor $\nu$ increases from below, $q_{\textrm{max}}$
abruptly rises from zero to a peak - whose magnitude depends on $t_{\textrm{am}}$
- once $\nu\apprge1$, and then gradually decreases. As a result,
we expect that the FFLO phase to be stabilized only at relatively
high filling factors, in stark contrast to the behavior found for
$d_{xy}$-wave altermagnetism.

\begin{figure}
\begin{centering}
\includegraphics[width=0.5\textwidth]{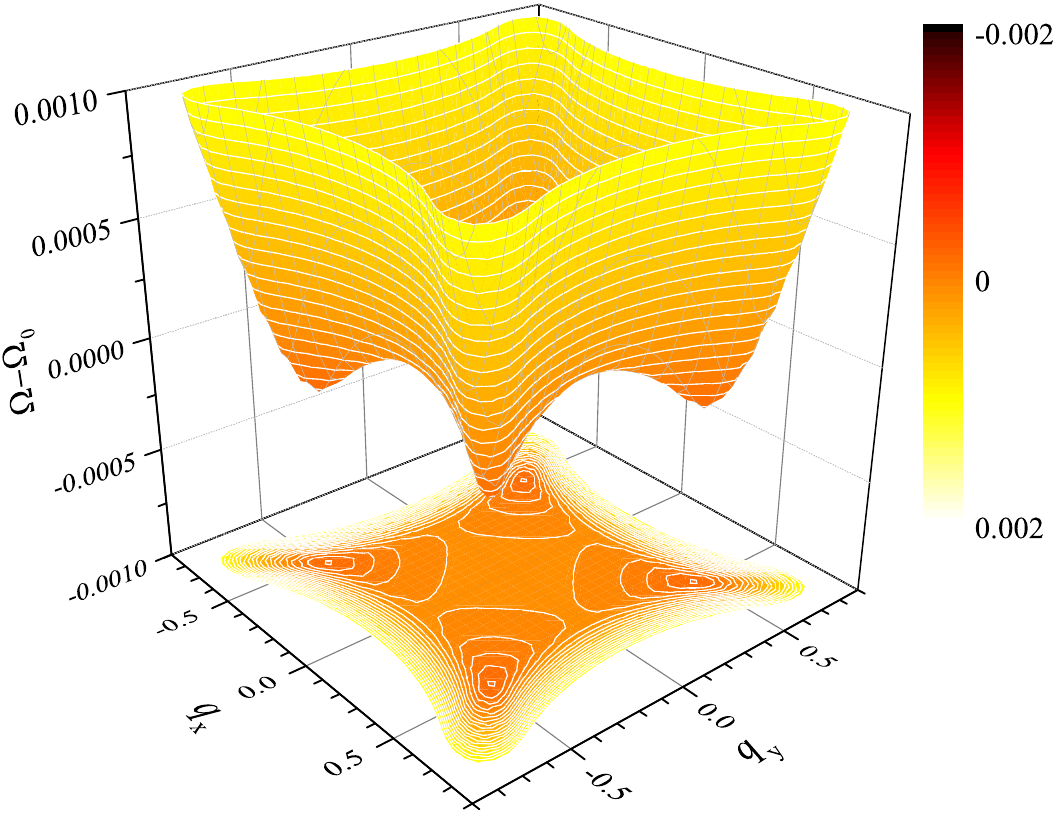}
\par\end{centering}
\caption{\label{fig7: Omega2dQXQY} The landscape of the thermodynamic potential
$(\Omega-\Omega_{0})/(\nu t)$, as a function of the FFLO momenta
$q_{x}$ and $q_{y}$, at the altermagnetic coupling strength $t_{\textrm{am}}=0.9t$
and the filling factor $\nu=1.2$. We have used a pairing gap $\Delta_{\textrm{FFLO}}=0.196t$
and a chemical potential $\mu_{\textrm{FFLO}}=2.62t$. The thermodynamic
potential reaches it minimum at $\mathbf{q}=(q_{x},q_{y})=(\pm q,\pm q)/\sqrt{2}$,
the diagonal or off-diagonal direction. Here, we consider the $d_{x^{2}-y^{2}}$-wave
altermagnetism and take an attraction strength $U=-3t$.}
\end{figure}

To verify such an anticipation, we carry out the Bogoliubov mean-field
calculations in the superconducting states. A representative result
is presented in Fig. \ref{fig7: Omega2dQXQY}, for the altermagnetic
coupling strength $t_{\textrm{am}}=0.9t$ and the filling factor $\nu=1.2$,
which yields a pairing amplitude $\Delta_{\textrm{FFLO}}\simeq0.196t$.
The resulting landscape of the thermodynamic potential $(\Omega-\Omega_{0})/(\nu t)$
over the $q_{x}-q_{y}$ plane clearly displays four minima located
along both the diagonal and off-diagonal directions. This agrees with
the non-self-consistent $T$-matrix analysis in the normal state,
which indicates that the pairing stability first sets in at $\mathbf{q}=(q_{x},q_{y})=(\pm q,\pm q)/\sqrt{2}$.

\begin{figure}
\begin{centering}
\includegraphics[width=0.5\textwidth]{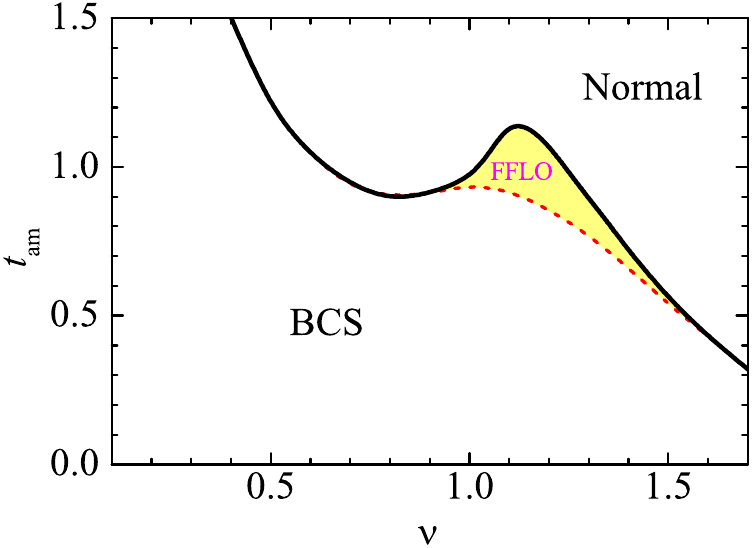}
\par\end{centering}
\caption{\label{fig8: dx2y2PhaseDiagram} The phase diagram under the $d_{x^{2}-y^{2}}$-wave
altermagnetism, as functions of the filling factor $\nu$ and the
altermagnetic coupling strength $t_{\textrm{am}}$. The transition
from the BCS state to the FFLO state is shown by the red dashed line,
while the transition from the FFLO state to the normal state is shown
by the black solid line. Here, we take an attraction strength $U=-3t$. }
\end{figure}

In Fig. \ref{fig8: dx2y2PhaseDiagram}, we report the phase diagram
as a function of the filling factor $\nu$ and the $d_{x^{2}-y^{2}}$-wave
altermagnetic coupling strength $t_{\textrm{am}}$. In sharp contrast
to the case of $d_{xy}$-wave altermagnetism, we observe that the
parameter window for the existence of the FFLO state is significantly
narrowed, indicating that $d_{x^{2}-y^{2}}$-wave altermagnetism is
much less effective to stabilizing FFLO superconductivity in lattice
systems \citep{Chakraborty2024}. For the transition from the superconducting
to normal state (i.e., black solid curve in the figure), we still
see a suppression of superconductivity around the filling factor $\nu=0.8$,
due to the presence of the Van Hove singularity.

\section{Conclusions and outlooks}

To conclude, we have investigated the inhomogeneous FFLO state induced
by altermagnetism in a two-dimensional metallic system, modeled by
electrons on lattices with different filling factors. Pairing instabilities
in the normal state were analyzed using a non-self-consistent $T$-matrix
approach, complemented by Bogoliubov mean-field calculations in the
superconducting phase. Our systematic investigation have revealed
that the parameter regime supporting the FFLO state is highly sensitive
to both the type of altermagnetism and the lattice filling. In the
case of $d_{xy}$-wave altermagnetism, the FFLO state is stabilized
over a broad range of filling factors. By contrast, for $d_{x^{2}-y^{2}}$-wave
altermagnetism, the FFLO phase is much more constrained, appearing
only at high filling and within a narrow parameter window at moderate
attractive interactions. This result is consistent with previous studies
showing that the FFLO state with $s$-wave interactions is difficult
to realize under $d_{x^{2}-y^{2}}$-wave altermagnetism \citep{Chakraborty2024}.

A crucial aspect of our study is the inclusion of next-to-nearest-neighbor
hopping in the single-particle dispersion, which serves as a simplified
yet effective way to emulate the band structure of real materials.
This additional hopping term breaks particle-hole symmetry and shifts
the location of the Van Hove singularity in the density of states
\citep{Romer2015}. We have found that the phase boundary associated
with the emergence of the FFLO state is strongly influenced by the
presence and position of the Van Hove singularity. This pronounced
effect is expected to persist even in more realistic two-band descriptions
of the single-particle dispersion that explicitly incorporate sub-lattice
degrees of freedom \citep{Sumita2025}.

In this work, we have focused on the simplest case of $s$-wave attractive
interactions, which may be mediated by phonons in an altermagnetic
metal. However, effective $d$-wave attractive interactions arising
from altermagnetic spin fluctuations are also a plausible pairing
mechanism. It would therefore be of considerable interest in future
studies to systematically explore altermagnetism-driven FFLO superconductivity
in a two-dimensional lattice at finite filling with high-partial-wave
pairing interactions.

\section{Statements and Declarations}

\subsubsection{Ethics approval and consent to participate }

Not Applicable.

\subsubsection{Consent for publication }

Not Applicable.

\subsubsection{Availability of data and materials }

The data generated during the current study are available from the
contributing author upon reasonable request. 

\subsubsection{Competing interests}

The authors have no competing interests to declare that are relevant
to the content of this article. 

\subsubsection{Funding}

This research was supported by the Australian Research Council's (ARC)
Discovery Program, Grants Nos. DP240100248 (X.-J.L.) and DP240101590
(H.H.). 

\subsubsection{Authors' contributions }

All the authors equally contributed to all aspects of the manuscript.
All the authors read and approved the final manuscript. 

\subsubsection{Acknowledgements }

See funding support.

\subsubsection{Authors' information}

Xia-Ji Liu, Centre for Quantum Technology Theory, Swinburne University
of Technology, Melbourne 3122, Australia, Email: xiajiliu@swin.edu.au

Hui Hu, Centre for Quantum Technology Theory, Swinburne University
of Technology, Melbourne 3122, Australia, Email: hhu@swin.edu.au

\appendix
%dummy comment inserted by tex2lyx to ensure that this paragraph is not empty

\begin{figure}
\begin{centering}
\includegraphics[width=0.45\textwidth]{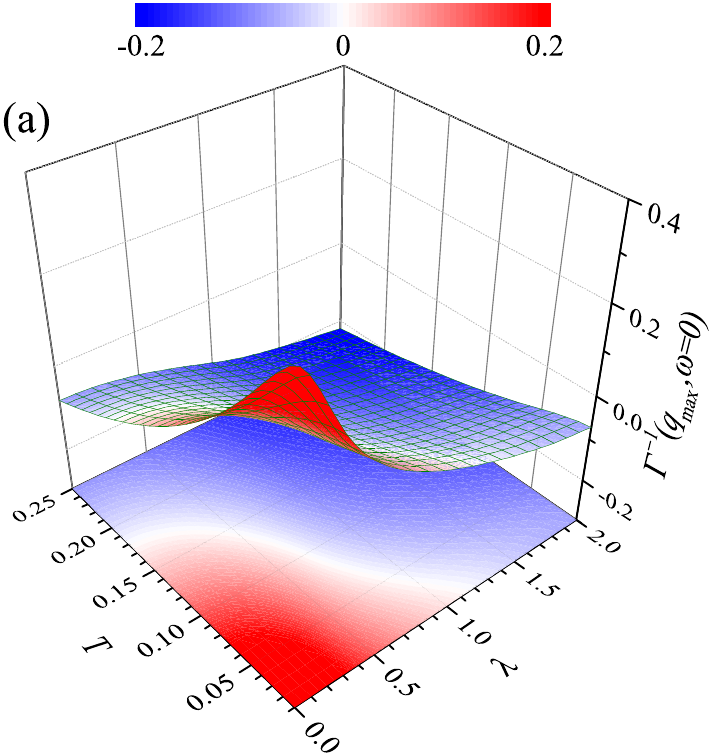}
\par\end{centering}
\begin{centering}
\includegraphics[width=0.5\textwidth]{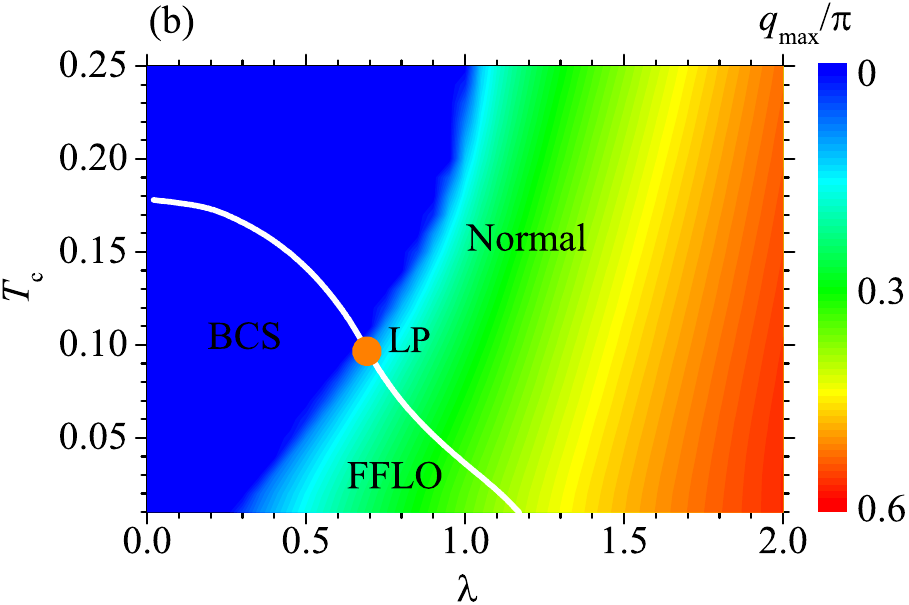}
\par\end{centering}
\caption{\label{fig9: tdepThouless} (a) The maximum of the inverse vertex
function $\Gamma^{-1}(q_{\max},\omega=0)$ and the corresponding pairing
momentum $q_{\max}$ (b) as functions of the altermagnetic coupling
$\lambda$ and the temperature $T$. The white line in (b) is the
critical temperature $T_{c}$ determined by the Thouless criterion
$\Gamma^{-1}=0$. The orange dot indicates the Lifshitz point at a
nonzero temperature. Here, we consider the $d_{xy}$-wave altermagnetism
and take an attraction strength $U=-3t$ at the filling factor $\nu=0.2$.}
\end{figure}

\section{The temperature effect}

Here, we examine the influence of finite temperature on altermagnetism-driven
FFLO superconductivity, focusing on the case of $d_{xy}$-wave altermagnetism
as a representative example. Figures \ref{fig9: tdepThouless}(a)
and \ref{fig9: tdepThouless}(b) display contour plots of $\Gamma^{-1}(q=q_{\textrm{max}},\omega=0)$
and of $q_{\textrm{max}}$, respectively, as functions of the altermagnetic
coupling strength $\lambda$ and temperature $T$. Throughout this
analysis, we fix the filling factor at a relatively low value, $\nu=0.2$.
As in previous figures, the superconducting region is highlighted
in red, while the transition to the normal state is delineated by
the white contour line.

As can be readily seen from Fig. \ref{fig9: tdepThouless}(b), the
pairing momentum $q_{\textrm{max}}$ decreases gradually with increasing
temperature. Consequently, the parameter regime supporting the uniform
BCS state (shown in blue) expands as the temperature rises. Overall,
the resulting finite-temperature phase diagram closely resembles that
obtained in the presence of an external magnetic field, which likewise
induces FFLO superconductivity at sufficiently large field strengths
\citep{Casalbuoni2004,Liu2006}. In particular, we identify a similar
Lifshitz point at a finite temperature $T_{c}\simeq0.56T_{c}^{(0)}$,
where $T_{c}^{(0)}\simeq0.18t$ denotes the BCS transition temperature
in the absence of altermagnetism.

\section{Altermagnetism-driven FFLO state in the limit of very low lattice
filling}

\begin{figure}
\begin{centering}
\includegraphics[width=0.5\textwidth]{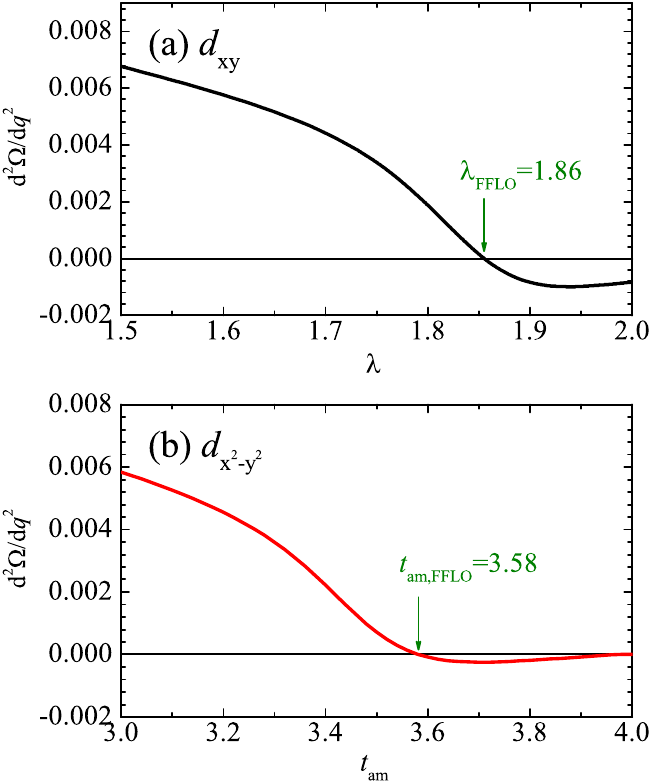}
\par\end{centering}
\caption{\label{fig10: d2Odq2_dilute} The second-order derivative $\partial^{2}\Omega/\partial q_{x}^{2}$
of the BCS state, in arbitrary units, as a function of the altermagnetic
coupling constant, in the case of $d_{xy}$-wave altermagnetism (a)
or $d_{x^{2}-y^{2}}$-wave altermagnetism (b). Here, we take an attraction
strength $U=-4t$ at a small filling factor $\nu=0.03$. The next-nearest-neighbor
hopping strength is set to be zero, $t'=0$.}
\end{figure}

\begin{figure*}
\begin{centering}
\includegraphics[width=0.45\textwidth]{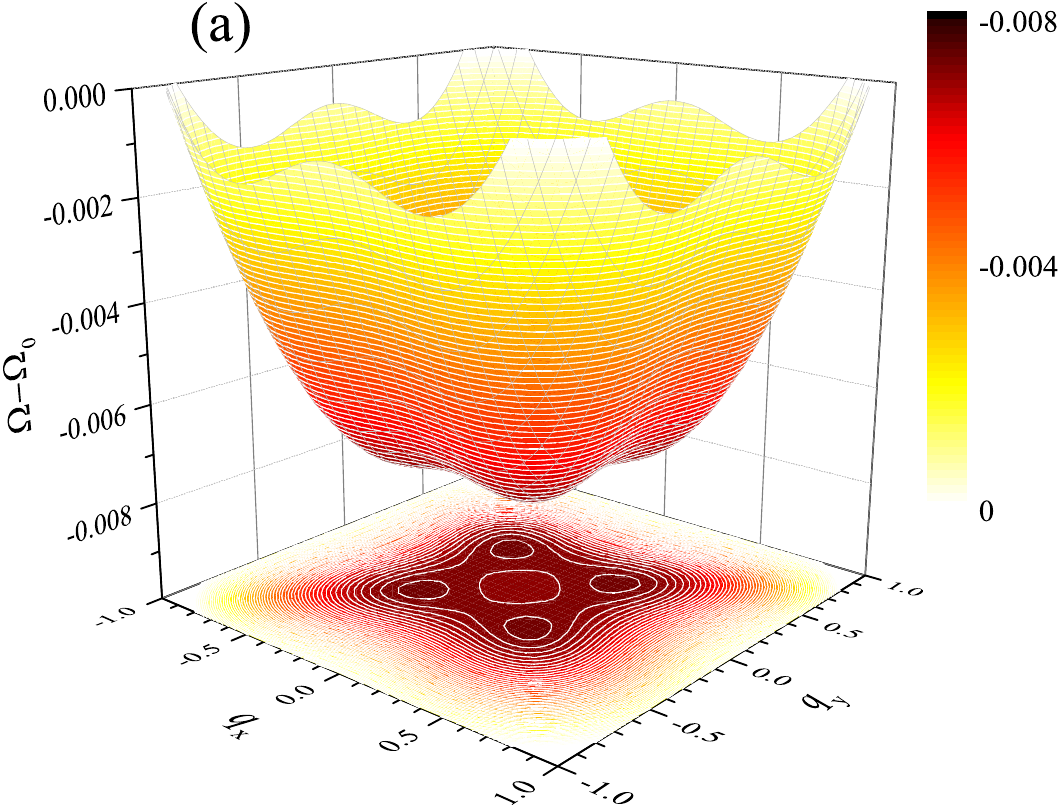}\includegraphics[width=0.45\textwidth]{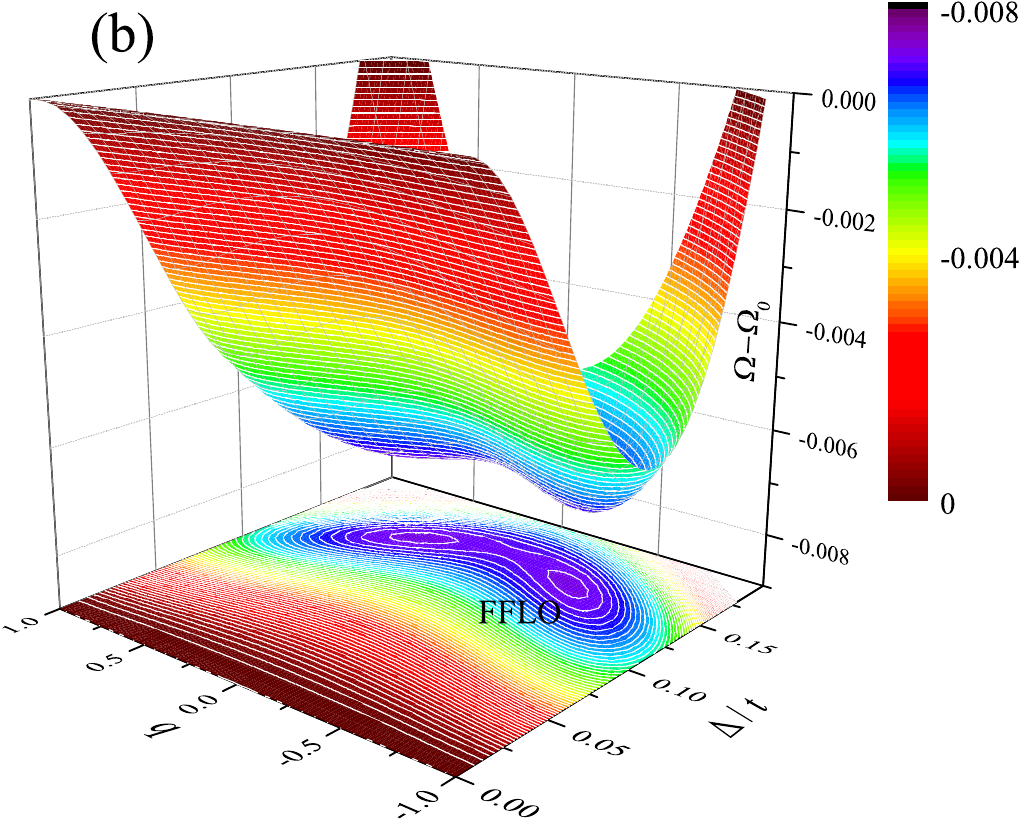}
\par\end{centering}
\caption{\label{fig11: dx2y2_dilute } (a) The landscape of the thermodynamic
potential $(\Omega-\Omega_{0})/(\nu t)$, as a function of the FFLO
momenta $q_{x}$ and $q_{y}$, at a pairing gap $\Delta_{\textrm{FFLO}}=0.149t$.
The thermodynamic potential reaches it minimum at $q_{x}=q_{y}=\pm q/\sqrt{2}$,
the diagonal or off-diagonal direction. (b) The landscape of $(\Omega-\Omega_{0})/(\nu t)$
as functions of the FFLO momenta $q=\pm\sqrt{q_{x}^{2}+q_{y}^{2}}$
and the pairing gap $\Delta$. In both plots, we have set the chemical
potential $\mu_{\textrm{FFLO}}=0.08t$. We consider the $d_{x^{2}-y^{2}}$-wave
altermagnetism with a coupling constant $t_{\textrm{am}}=3.7t$, and
take an attraction strength $U=-4t$. The filling factor of the lattice
is $\nu=0.03$.}
\end{figure*}

Earlier studies on the altermagnetism-driven FFLO state in 2D lattices
have primarily focused on the regime of low filling \citep{Chakraborty2024,Hong2025,Hu2025PRB},
where a continuum model Hamiltonian can be employed as an effective
description \citep{Hu2025PRB,Liu2025PRB,Soto-Garrido2014}. In this
dilute limit, the specific type of $d$-wave altermagnetism becomes
irrelevant. To illustrate this point, we expand the non-interacting
single-particle dispersion relation, including altermagnetic effects,
for small $k_{x}$ and $k_{y}$ as follows,

\begin{equation}
\xi_{\uparrow,\downarrow}^{(xy)}\left(\mathbf{k}\right)=-4t+t\left(k_{x}^{2}+k_{y}^{2}\right)\pm\lambda k_{x}k_{y}
\end{equation}
for $d_{xy}$-wave altermagnetism, and 
\begin{equation}
\xi_{\uparrow,\downarrow}^{(x^{2}-y^{2})}\left(\mathbf{k}\right)=-4t+t\left(k_{x}^{2}+k_{y}^{2}\right)\pm\frac{t_{\textrm{am}}}{4}\left(k_{x}^{2}-k_{y}^{2}\right)
\end{equation}
for $d_{x^{2}-y^{2}}$-wave altermagnetism. Here, for simplicity we
have set the next-to-nearest-neighbor hopping strength $t'=0$. It
is straightforward to verify that, upon performing a $\pi/4$ rotation
in momentum space, i.e., $k_{x}\rightarrow(k_{x}+k_{y})/\sqrt{2}$
and $k_{y}\rightarrow(-k_{x}+k_{y})/\sqrt{2}$, the dispersion $\xi_{\uparrow,\downarrow}^{(xy)}(\mathbf{k})$
acquires the same functional form as $\xi_{\uparrow,\downarrow}^{(x^{2}-y^{2})}(\mathbf{k})$,
provided that the parameters are identified via, 
\begin{equation}
\lambda\longleftrightarrow\frac{t_{\textrm{am}}}{2}.
\end{equation}

In Fig. $\ref{fig10: d2Odq2_dilute}$, we present the phase stiffness
$\rho_{s}\sim\partial^{2}\Omega/\partial q_{x}^{2}$ of the uniform
BCS state as a function of increasing altermagnetic coupling. Results
are shown for the $d_{xy}$-wave altermagnetism (upper panel) and
the $d_{x^{2}-y^{2}}$-wave altermagnetism (lower panel), considering
a relatively strong attractive interaction $U=-4t$ and a low filling
factor $\nu=0.03$. Consistent with expectations, the threshold altermagnetic
coupling strengths in the two cases approximately obey the relation
$\lambda_{\textrm{FFLO}}=t_{\textrm{am},\textrm{FFLO}}/2$. 

To conclusively demonstrate that the FFLO state emerges under $d_{x^{2}-y^{2}}$-wave
altermagnetism in the dilute limit, we show in Fig. \ref{fig11: dx2y2_dilute }
the landscape of the thermodynamic potential $(\Omega-\Omega_{0})/(\nu t)$
at a large altermagnetic coupling strength $t_{\textrm{am}}=3.7t$.
The FFLO phase is clearly identified as the global minimum of the
thermodynamic potential landscape.

\end{document}